\documentclass[nofootinbib,superscriptaddress,reprint,onecolumn,amsmath,amssymb,mathtools,booktabs,array,10pt,notitlepage,aps]{revtex4-1}

\usepackage{graphicx}
\usepackage[T1]{fontenc}
\usepackage{epstopdf}
\usepackage{dcolumn}
\usepackage{bm}

\usepackage{hyperref}
\hypersetup{colorlinks, linkcolor={DarkRed},citecolor={NavyBlue},urlcolor={NavyBlue}}

\usepackage[dvipsnames, svgnames, x11names]{xcolor}
\usepackage{float}
 \usepackage[mathscr]{eucal}
\def\d{{\mathrm d}}

\usepackage{mathtools}
\usepackage{textcomp}

\begin{document}

\title{Bounce corrections to gravitational lensing, quasinormal spectral stability and gray-body factors of  Reissner-Nordstr\"om black holes}

\author{Yang Guo}\email{ guoy@mail.nankai.edu.cn}
\affiliation{School of Physics, Nankai University, Tianjin 300071, China}
\author{Chen Lan}\email{lanchen@nankai.edu.cn}
\affiliation{Department of Physics, Yantai University, Yantai 264005, China},
\affiliation{School of Physics, Nankai University, Tianjin 300071, China}
\author{Yan-Gang Miao}\email{Corresponding author: miaoyg@nankai.edu.cn}
\affiliation{School of Physics, Nankai University, Tianjin 300071, China}


\begin{abstract}
Gravitational lensing in the weak field limit, quasinormal spectra, and gray-body factors are investigated in the Reissner-Nordstr\"om spacetime corrected by bounce parameters. Using the Gauss-Bonnet theorem, we analyze the effects of bounce corrections to the weak gravitational deflection angle and find that the divergence of the deflection angle can be suppressed by a bounce correction in the Reissner-Nordstr\"om spacetime. We also notice that the bounce correction plays the same role as the Morse potential in the deflection angle. Moreover, we derive the perturbation equations with spin-dependent effective potentials and discuss the quasinormal spectral stability. We observe that the quasinormal spectra have the same behavior in large bounce parameters for different multiple
numbers and they decrease significantly if the bounce parameter is much bigger than the charge. We further study the transmission probability of particles scattered by the effective potentials and reveal that the bounce correction introduced into the Reissner-Nordstr\"om spacetime increases the gray-body factors of perturbation fields.

\end{abstract}

\maketitle
\section{Introduction}
\label{sec:intro}
The singularity problem of a spacetime has always been a topic of great concerns in general relativity (GR) and black hole physics. The singularity theorem established  by Hawking and Penrose \cite{hawking1973large} claims that the singularities are an inevitable feature of Einstein's theory. However, it is commonly believed that such singularities are indeed nonphysical objects occurred in classical theories of gravity and the occurrence of singularities is considered to be an indicator that GR should be modified and generalized to a quantum theory. Following the early quantum arguments of Sakharov \cite{sakharov1966initial} and Gliner \cite{gliner1966algebraic} that the singularities could be avoided by the quantum influence of matter sources, i.e., replacing the black hole singularity with a de Sitter core, Bardeen and others proposed \cite{bardeen1968proceedings,Kazakov:1993ha,Borde:1996df,Dymnikova:2001fb,Hayward:2005gi,Frolov:2017rjz} various modifications  of the Schwarzschild black hole, see, for instance, some comprehensive review articles \cite{Ansoldi:2008jw,Frolov:2014jva} on  regular black holes. In addition, some excellent and  lively arguments have been suggested \cite{Fan:2016hvf,Bronnikov:2017tnz,Toshmatov:2018cks,Bonanno:2020fgp}  in the construction of regular black holes in GR.  

More recently, there has been a renewed interest for the search of alternatives to classical black holes in GR. The research originates from a bounce parameter associated with the Plank scale  introduced  by Simpson and Visser \cite{Simpson:2018tsi} in the modification of the Schwarzschild black hole.  A great variety of solutions based on bounce and quantum corrections have been obtained \cite{Berry:2021hos,Huang:2019arj,Franzin:2021vnj,Mazza:2021rgq,Xu:2021lff}, which provides us with the treatment for the singularities of black holes. All of these black hole mimickers are globally free from curvature singularities. Especially, the black-bounce family  passes all weak-field observational tests, and it smoothly interpolates regular black holes and  traversable wormholes. In this paper, we focus our attention on the bounce corrections at the interface between the Reissner-Nordstr\"om black hole and a regular black hole.

It is well known that GR describes how matter distorts the spacetime around it. The gravitational lensing occurs when a huge amount of matter creates a gravitational field distorting the light from a source. As a significant phenomenon, the gravitational lensing can reflect the distribution of matter, such as galaxy clusters \cite{Hoekstra:2013via,Brouwer:2018xnj,Bellagamba:2018gec,Hoekstra:2003pn}, dark matter \cite{Jung:2017flg,Andrade:2019wzn,Turimov:2018ttf}, dark energy \cite{Vanderveld:2012ec,Cao:2012ja,Huterer:2017buf}, black holes \cite{Bhadra:2003zs,Chen:2009eu,Eiroa:2002mk,Virbhadra:1999nm,Okyay:2021nnh,Guo:2021wid,Zhang:2021ygh}, and wormholes \cite{Nandi:2006ds,Virbhadra:2002ju,Sharif:2015qfa,Ovgun:2020yuv}, etc.   
Gibbons and Werner applied~\cite{Gibbons:2008rj} the Gauss-Bonnet theorem to develop an alternative approach with a global feature to gravitational lensing theories. With the help of this feature, we consider the weak deflection limit and treat light rays as spatial geodesics of the optical geometry. One of our main aims is to clarify the effects of bounce parameters on the gravitational deflection angle in the Reissner-Nordstr\"om black hole.

Additionally,  quasinormal modes (QNMs) have been studied \cite{Berti:2003ud,Guo:2020nci,Bronnikov:2021liv,Guo:2020caw,Konoplya:2019xmn} in a wide range of issues in the context of GR and alternative theories of gravity.
QNMs are usually used to depict the stability of black holes perturbed by an external field, and  also contain the information of gravitational waves. The fundamental mode is the least damped and long lived mode in a ringdown signal and is more likely to be used to test the (in)stability of black holes. On the other hand, the gray-body factors encode \cite{Kanti:2002nr} information about the horizon structure of black holes theoretically and modify the quasinormal spectra experimentally. For estimating effectively the transmission probability of radiations from a black hole's event horizon to its asymptotic region, we need to investigate the gray-body factors of perturbations. We derive the perturbation field equation with a spin-dependent effective potential and determine the quasinormal spectra numerically. 
Moreover, we calculate the gray-body factors of waves scattered by effective potentials.
Here we focus on the effects of bounce parameters  introduced in the Reissner-Nordstr\"om black hole.

The outline of this paper is as follows. In Sec.~\ref{sec:thermodynamics} we review briefly the  properties of the black-bounce-Reissner-Nordstr\"om geometry and describe the main aspects of bounce corrections. Next we apply  in Sec.~\ref{sec:deflection angle} the Gauss-Bonnet theorem to the gravitational lensing in the weak field limit and investigate the gravitational deflection angle corrected by bounce parameters.  We derive the master wave equation with a spin-dependent effective potential under the massless scalar and electromagnetic field perturbations in Sec.~\ref{sec:RWanalysis}. We then calculate the quasinormal spectra and discuss the spectral stability in Sec.~\ref{sec:QNM}. The bounce corrections to the gray-body factors of perturbation fields are computed in Sec.~\ref{sec:GBF}.   Finally, we give our conclusions in Sec.~\ref{sec:con}.	
Appendices \ref{app:recurrence} and \ref{app:gaussian} include the detailed calculations of quasinormal spectra in terms of the Leaver's method. 

\section{Reissner-Nordstr\"om geometry corrected by bounce parameters}
\label{sec:thermodynamics}

A regularizing procedure has recently been introduced \cite{Franzin:2021vnj} into Reissner-Nordstr\"om black holes, which does not generate \cite{Ayon-Beato:2000mjt,Hayward:2005gi,Frolov:2017rjz} a traditional regular black hole, such as the Bardeen's or Hayward's.
This procedure gives rise to a one-parameter modification of the  Reissner-Nordst\"om
black hole of general relativity, and can be obtained as an exact solution to the Einstein equations sourced by a combination of a minimally coupled phantom scalar field  and a  nonlinear electrodynamics field. The action reads~\cite{Bronnikov:2021uta},
\begin{eqnarray}
	S=\int\sqrt{-g}\,d^4x \left( \mathcal{R} + 2 \epsilon g^{\mu\nu}\partial_\mu\phi \partial_\nu\phi - 2V(\phi) 
	- \mathcal{L(F)} \right),\label{action}
\end{eqnarray}
where $\mathcal{L(F)}$ is the Lagrangian density of gauge-invariant nonlinear electrodynamics  with $\mathcal{F}\equiv F_{\mu\nu}F^{\mu\nu}$, and  $\epsilon=-1$ for a phantom scalar field.  The Lagrangian density and the potential of a phantom scalar field $\phi(x)$ take the following forms,
\begin{eqnarray}
	\mathcal{L(F)} 	= \frac{12 ma^2}{5 (2q^2/\mathcal{F})^{5/4}} 
	+ \frac {2Q^2 \big[3(2q^2/\mathcal{F})^{1/2}-4a^2\big]}{3(2q^2/\mathcal{F})^{3/2}},	\label{V-fin}   
\end{eqnarray}
and
\begin{eqnarray}
	V(\phi) = \frac {2 \cos^6 \phi}{15 a^4} (6m a \sec \phi - 5Q^2),
	\label{V-phant}
\end{eqnarray}
where $m$ is mass of the black-bounce, $q$ magnetic charge of free nonlinear electrodynamics, $Q$ electric charge parameter, and $a$ the bounce parameter of the charged black-bounce, respectively.

Varying Eq.~(\ref{action}) with respect to the metric yields Einstein's equations,
\begin{eqnarray}
	G_{\mu}^{\nu}=-T_{\mu}^{\nu},
	\label{Einequ}
\end{eqnarray}
and varying the action with respect to $\phi$ and $F_{\mu\nu}$, respectively, gives the field equations,
\begin{eqnarray}
	2\epsilon\nabla_\mu\nabla^\mu\phi+\frac{dV(\phi)}{d\phi}=0,
	\label{Phantequ}
\end{eqnarray}
and
\begin{eqnarray}
	\nabla_\mu(\mathcal{L(F)}F^{\mu\nu})=0.
	\label{Nonlinearequ}
\end{eqnarray}
The  stress-energy tensor is a combination of the stress-energy tensor of the scalar field  and that of the nonlinear electromagnetic field,
\begin{eqnarray}  T_{\mu}^{\nu}=T_{\mu}^{\nu}[\phi]+T_{\mu}^{\nu}[{F}],
	\label{Totstrerg} 
\end{eqnarray}	
where $T_{\mu}^{\nu}[\phi]$ and $T_{\mu}^{\nu}[{F}]$ take the forms,
\begin{eqnarray}
	T_{\mu}^{\nu}[\phi]=2\epsilon\partial_\mu\phi\partial^\nu\phi- \delta_{\mu}^{\nu}(\epsilon g^{\rho\sigma}\partial_\rho\phi\partial_\sigma\phi- V(\phi)),
	\label{Phantstrerg}
\end{eqnarray}
and
\begin{eqnarray}
	T_{\mu}^{\nu}[{F}]=-2\frac{d\mathcal{L}}{d\mathcal{F}}F_{\mu\rho}F^{\nu\rho}+\frac{1}{2}\delta_\mu^\nu\mathcal{L}(\mathcal{F}).
	\label{Nonlinearstrerg}
\end{eqnarray}
Here the phantom scalar field, $\phi(x)$, reads 
\begin{eqnarray}
	\phi(x)=\pm \tan^{-1}\frac{x}{a}+{\rm const}.
	\label{phantfield}
\end{eqnarray}
Now we can write the non-zero components of the stress-energy tensor by using Eqs. (\ref{V-fin}), (\ref{V-phant}), and (\ref{Totstrerg})-(\ref{phantfield}),
\begin{eqnarray}
	T_t^t=-\frac{a^4+a^2 \left(-4 m \sqrt{a^2+r^2}+2 Q^2+r^2\right)-Q^2 r^2}{\left(a^2+r^2\right)^3},\label{ttcomp}
\end{eqnarray}
\begin{eqnarray}
	T_r^r=\frac{a^4+r^2 \left(a^2+Q^2\right)}{\left(a^2+r^2\right)^3},\label{rrcomp}
\end{eqnarray}
\begin{eqnarray}
	T_\theta^\theta=T_\varphi^\varphi=-\frac{a^4+a^2 \left(r^2-m \sqrt{a^2+r^2}\right)+Q^2 r^2}{\left(a^2+r^2\right)^3}.\label{thetacomp}
\end{eqnarray}
In a four dimensional static and spherically symmetric spacetime, by considering the following line element, 
\begin{eqnarray}
	ds^2=-f(r)dt^2+f^{-1}(r)dr^2+h^2(r)d\Omega^2, \label{bbrn}
\end{eqnarray} 
and substituting it into $G_{\mu}^{\nu}$, we can find the shape function by solving Eqs. (\ref{Einequ})-(\ref{Nonlinearequ}) with the aid of Eqs. (\ref{ttcomp})-(\ref{thetacomp}),
\begin{eqnarray}
	f(r)=1-\frac{2m}{\sqrt{r^2+a^2}}+\frac{Q^2}{r^2+a^2}, \qquad h^2(r)\equiv r^2+a^2.\label{shape}
\end{eqnarray} 

This one-parameter modification of Reissner-Nordstr\"om black holes, Eqs.~(\ref{bbrn}) and (\ref{shape}), is also called a {\em black-bounce-Reissner-Nordstr\"om}, {\em charged black-bounce}, or {\em Reissner-Nordst\"om-Simpson-Visser} (RN-SV) spacetime which gives either a charged regular black hole or a traversable wormhole. Whether it is a regular black hole or a wormhole depends on the value of the bounce parameter $a$.  Due to the high degree of mathematical tractability,  this class of geometries is labeled as "bounce".
The introduction of bounce parameters was partly inspired and motivated by Kazakov and Solodukhin's work~\cite{Kazakov:1993ha} on the quantum deformed Schwarzschild spacetime. The quantum corrections to theories of gravity may completely change the gravitational equations and the corresponding  geometry at the Planck scale. This will lead to a deformation due to quantum excitations of the metric and matter fields. The RN-SV spacetime proposed by Simpson and Visser~\cite{Simpson:2018tsi} is based on Kazakov and Solodukhin's results and the parameter $a$ in the RN-SV spacetime is often identified with a deformation parameter related to the Planck scale.

Several properties of the black-bounce family have been well tested \cite{Lobo:2020ffi,Simpson:2019cer,Chakrabarti:2021gqa}: (i) The black-bounce family is globally free from curvature singularities; (ii) It passes all weak-field observational tests. We note that the radial coordinate expands to the entire real domain, $r\in(-\infty, +\infty)$, so a  coordinate speed of light can be defined~\cite{Simpson:2018tsi,Bronnikov:2021liv}  in terms of the radial null curves ($\d s^2=0$ and $\d\theta=\d\varphi=0$), 
\begin{eqnarray}
   c(r)=\left|\frac{\d r}{\d t}\right|=1-\frac{2m}{\sqrt{r^2+a^2}} +\frac{Q^2}{r^2+a^2},
\end{eqnarray}
and the area of a sphere at radial coordinate $r$ takes the following form in this spacetime,
\begin{eqnarray}
	A(r)=4\pi h^2(r).
\end{eqnarray}
The area is minimized at the wormhole throat and one can find the location of the throat by the condition,
\begin{eqnarray}
	 A^{\prime}(r_0)=0,
\end{eqnarray}
where $r_0$ is the location of the throat. Then $h_0\equiv h(r_0)$ corresponds to the radius of the wormhole throat. Now we classify this geometry into three types: 
\begin{itemize}
	\item {$a<m\pm\sqrt{m^2-Q^2}$ and $|Q|<m$,  there exists one outer/inner horizon at $r_{\rm h}=\pm\sqrt{( m\pm\sqrt{m^2-Q^2}) ^2-a^2}$. In this case, we obtain
	\begin{eqnarray}
		\exists\,  r_{\rm h}\in\mathbb{R}^*: \quad c(r_{\rm h})=0.
\end{eqnarray}
The coordinate speed of light is zero and the light cannot escape from the horizon. This geometry is clearly a charged regular black hole with a standard outer/inner horizon.}
	\item  {$a=m\pm\sqrt{m^2-Q^2}$ and $|Q|<m$, there exists one extremal horizon at $r_{\rm h}=0$. Hence, we know
	\begin{eqnarray}
		\exists\, r_{\rm h}=0: \quad c(r_{\rm h})=0.
\end{eqnarray} The geometry corresponds to one extremal charged regular black hole. Alternatively, it is called a one-way charged traversable wormhole with one extremal null throat located at $r_0=0$. }
	\item  {$a>m\pm\sqrt{m^2-Q^2}$ and $|Q|<m$ or $|Q|>m$, there exist no horizons. We have
		\begin{eqnarray}
			\forall\, r\in(-\infty, +\infty): \quad c(r)\neq0.
		\end{eqnarray} The light can travel across the entire domain. So this geometry is a two-way charged traversable wormhole with the radius, $h=a$. }
\end{itemize}

\section{ Gravitational lensing in the weak field limit} 
\label{sec:deflection angle}
In this section, we present an investigation to the gravitational lensing in a charged black-bounce using the Gauss-Bonnet theorem. The fact that the Gauss-Bonnet theorem can be used for characterising  lensing features has been demonstrated \cite{Gibbons:2008rj} by Gibbons and Werner  who applied the Gauss-Bonnet theorem to a static, spherically symmetric, and perfect non-relativistic fluid in the weak deflection limit as a simple model of gravitational lens. The intrinsic geometric and topological properties of a surface are linked by the Gauss-Bonnet theorem,
  \begin{eqnarray}
	\iint_D K\,\d S + \int_{\partial D} \kappa\,\d t + \sum_{i} \alpha_i = 2\pi \chi(D).\label{eq:GBT}
\end{eqnarray}
Here the first term represents the integral of Gaussian curvature $K$ over a compact oriented surface $D$ with Euler characteristic number $\chi(D)$. The second term is the integral of geodesic curvature $\kappa$ over the boundary of $D$, and $\alpha_i$ denotes the  exterior angle at the $i$th vertex. In the center of lens without singularity the Euler characteristic number equals one, $\chi(D)=1$.  The optical metric of the charged black-bounce can be derived from the null geodesic, $\d s^2=0$, in the equatorial plane $(\theta=\pi/2)$,
\begin{eqnarray}
	\d t^2=\frac{\d r^2}{\left( 1-\frac{2m}{\sqrt{r^2+a^2}} +\frac{Q^2}{r^2+a^2}\right) ^2} + \frac{(r^2+a^2)\d\varphi^2}{ 1-\frac{2m}{\sqrt{r^2+a^2}} +\frac{Q^2}{r^2+a^2}}.
\end{eqnarray}
 The Gaussian curvature of the optical metric can be calculated,
 \begin{eqnarray}
 	K& \approx&-\frac{2m}{r^3}+\frac{-a^2+3 m^2+3 Q^2}{r^4}+\frac{10m a^2 -6 m Q^2}{r^5}+\frac{2 a^4-15 a^2 m^2-12 a^2 Q^2+2 Q^4}{r^6}\nonumber \\  
 	& &+\frac{-85m a^4+112m a^2  Q^2 }{4 r^7} +\frac{36m^2 a^4 +27 a^4 Q^2-10 a^2 Q^4}{r^8}-\frac{287 ma^4  Q^2}{4 r^9}+\frac{28 a^4 Q^4}{r^{10}} + \mathcal{O}(a^5).
 \end{eqnarray}

 The specific domain $D$ denoting the weak deflection lensing geometry is bounded by a circular curve $C_R$ and a geodesic $\gamma$ from the source to an observer. So its boundary $\partial D$ consists of two parts: The geodesic $\gamma$ and a circular curve $C_R$. We note that the geodesic curvature along the geodesics vanishes, i.e., $\kappa(\gamma)=0$. 
 If the source and observer are located at an infinite distance from the lens, for the circular curve $C_R :=r(\varphi)=R=\text {const.}$,  the geodesic curvature can be defined as
 \begin{eqnarray}
 	\kappa(C_R)=|\nabla_{\dot{C}_R}\dot{C}_R|,
 \end{eqnarray}
 where $\dot{C}_R$ is the tangent vector of $C_R$.  The integral over the boundary can be reduced to be
 \begin{eqnarray}
 	\int_{\partial D} \kappa \,\d t&=&\int_{\gamma} \kappa(\gamma) \, \d t + \int_{C_R} \kappa(C_R)\,\d t \nonumber\\
 	&=&\lim_{R\to\infty}\int_{C_R}\kappa(C_R)\,\d t.
 \end{eqnarray}
 In the limit of $R\to \infty$, we have $\kappa(C_R)\d t=\lim_{R \to \infty}[\kappa(C_{R})\d t]=\d\varphi$. The Gauss-Bonnet theorem can be rewritten as
 \begin{eqnarray}
 		\iint_D K\,\d S + \int_{0}^{\pi+\alpha} \,\d\varphi  = \pi .
 \end{eqnarray}

\begin{figure}[htbp]
	\centering
	\includegraphics[width=0.48\linewidth]{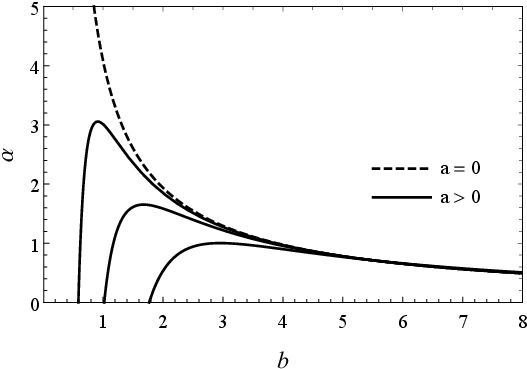}\includegraphics[width=0.512\linewidth]{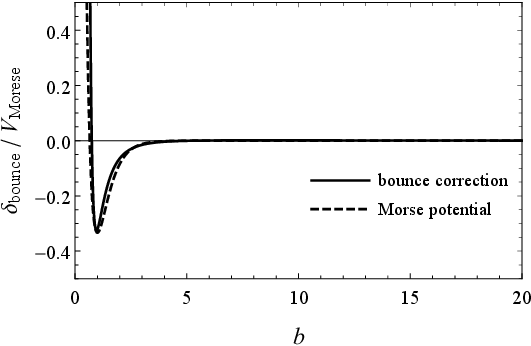}
	\caption{The deflection angle with respect to the impact parameter in the weak field limit (left), and the bounce parameter correction with respect to the impact parameter  in the charged black-bounce spacetime (right), where $m=1$, $Q=1/2$, and $a=1/2$ are set. In the left panel, the black dashed curve with a zero bounce correction ($a=0$) corresponds to the deflection angle of the Reissner-Nordstr\"om black hole. In the right panel, the solid black curve corresponds to the bounce correction with $m=1$, $Q=1/2$, and $a=1/2$, and the Morse potential (dashed black curve) is chosen for comparison when  $x_0=1$, $\tilde D=1/3$, and $\tilde\gamma=2$ are set.}
	\label{fig:angle}
\end{figure}

In the weak field deflection limit,  the zeroth order light ray with impact parameter $b$ is given by $r(t)=b/\sin\varphi$. Therefore, the weak gravitational deflection angle for the charged black-bounce can be determined by 
 \begin{eqnarray}\label{angle}
	\alpha_{\text{charged black-bounce}} & =& - \iint_D K \, \d S=-\int_{0}^{\pi}\int_{b/\sin\varphi}^{\infty}K\,\d S \nonumber\\
	&\approx&\underbrace{\frac{4 m}{b}}_{\alpha_{\text{Schwarzschild}}} \underbrace{- \frac{3\pi Q^2}{4 b^2} +\frac{8 m  Q^2}{3 b^3}}_{\delta_{\text{electrodynamics}}} +\overbrace{\frac{\pi a^2}{4b^2}-\frac{40ma^2}{9b^3}+ \frac{9 \pi   a^2  Q^2}{8 b^4}- \frac{ 448ma^2 Q^2}{75 b^5}}^{\delta_{\text{bounce}}} + \mathcal{O}(m^2,a^4,Q^4).
\end{eqnarray}
The leading order in Eq. (\ref{angle}) is known \cite{Gibbons:2008rj} as the deflection angle for the Schwarzschild black hole. The second and third terms are the contributions  \cite{Jusufi:2015laa,Fu:2021akc}  from the pure electric sources. It is clear that there exist extra correction terms associated with the bounce parameter labeled by $\delta_{\rm bounce}$. In general, for a traditional black hole the deflection angle increases \cite{Javed:2021arr,Jusufi:2017mav} continuously with the decrease of the impact parameter $b$ and it eventually diverges. The deflection angle for the Reissner-Nordstr\"om black hole as a special case (corresponding to $a=0$ in the charged black-bounce) is shown in the left panel of Fig. \ref{fig:angle}. However, for $a>0$, we can observe that the deflection angle is finite due to the bounce correction when the impact parameter reduces. The deflection angle is suppressed by the bounce correction $\delta_{\text{bounce}}$, which is similar in shape to the Morse potential,\footnote{The Morse potential \cite{Morse:1929zz},  proposed by Phillip M. Morse in 1929,  describes an interaction model that consists of diatomic molecules. It has the form, $V_{\text{Morse}}=\tilde D\left(1-e^{-\tilde\gamma x}\right)^2$ or $V_{\text{Morse}}=\tilde D\left[e^{-2\tilde\gamma (x-x_0)}-2e^{-\tilde\gamma (x-x_0)}\right]$. For a comprehensive introduction and a recent application to quasinormal spectral problems, see Ref. \cite{costa2013morse} and Ref. \cite{Hatsuda:2021gtn}, respectively.}
\begin{eqnarray}
	V_{\text{Morse}}(x)=\tilde D\left[e^{-2\tilde\gamma (x-x_0)}-2e^{-\tilde\gamma (x-x_0)}\right],
\end{eqnarray}
where $x$ is the distance between atoms, $x_0$ is the location of the minimum potential, $\tilde D$ is the well depth, and $\tilde\gamma$ is a length parameter related to the width of the well. We find that the bounce correction to the deflection angle for the Reissner-Nordstr\"om black hole has the same form as the Morse potential if we identify the distance between atoms with the impact parameter, which is shown in the right panel of Fig. \ref{fig:angle}.
On the other hand, since the Morse potential is asymptotically flat (corresponding to $\delta_{\text{bounce}}$ asymptotically vanishing), the deflection angle for different bounce parameters is almost the same when $b$ increases. That is, the effect of increasing $b$ on the deflection angle is negligible. Finally, we point out that the deflection angle for the charged black-bounce is composed of two parts, one is the Reissner-Nordstr\"om deflection angle and the other is the bounce correction term,
\begin{eqnarray}
	\alpha_{\text{charged black-bounce}} = \alpha_{\text{Reissner-Nordstr\"om}} + \delta_{\text{bounce}}.
\end{eqnarray}

\section{The master wave equations for scalar and electromagnetic perturbations}
\label{sec:RWanalysis}

In a spherically symmetric background, the evolution of linearized perturbation fields of spin $s$ is described \cite{Berti:2009kk,Konoplya:2011qq,Kokkotas:1999bd} by the master wave equation,
\begin{eqnarray}
		\frac{\d ^2\Psi_s}{\d r_*^2}+(\omega^2-V_s)\Psi_s=0, \label{masterEq}
\end{eqnarray}
where $r_*$ is ``tortoise'' coordinate defined by the relation, $\d r_*/\d r=1/f(r)$. To simplify the notation of the equation, we have taken the $s$-subscript $\Psi_s$ and $V_s$, where $\Psi_s$ denotes the scalar or vector field oscillating and decaying at a complex frequency $\omega$, and $V_s$ is the spin-dependent effective potential. 

Consider a massless scalar ($s=0$) perturbation field propagating in a curved spacetime,  its wave equation satisfies
\begin{eqnarray}
	\frac{1}{\sqrt{-g}}\partial_{\mu} \left(\sqrt{-g}g^{\mu \nu}\partial_{\nu} \phi \right)=0, \label{KGEq}
\end{eqnarray}
where $g$ and $g^{\mu\nu}$ denote the determinant and  inverse of $g_{\mu\nu}$, respectively. In the spacetime equipped with  a time-independent and spherically symmetric metric,  Eqs.~(\ref{bbrn}) and (\ref{shape}), we can decompose $\phi (t,r,\theta,\varphi)$ into Fourier modes,
\begin{eqnarray}
	\phi(t,r,\theta,\varphi)=\sum_{\ell,m}e^{-iwt} \frac{\Psi_{s=0}(r)}{h(r)} Y_{\ell m}(\theta,\varphi),\label{decompose}
\end{eqnarray} 
 and redefine $\Psi_{s=0}(r)$ as the perturbation field, where $Y_{\ell m}(\theta,\varphi)$ stands for the spherical harmonics. Substituting the decomposition Eq.~(\ref{decompose}) into Eq. (\ref{KGEq}), we can get the master equation Eq.~(\ref{masterEq}) for $\Psi_{s=0}(r)$ with the effective potential,
\begin{eqnarray}
	V_{s=0}=f(r)\left\lbrace \frac{\ell(\ell+1)}{h^2(r)} +\frac{1}{h(r)}\frac{\d }{\d r}\left[ f(r)\frac{\d h(r)}{\d r}\right]  \right\rbrace .
\end{eqnarray}

For a  linearized Maxwell ($s=1$) field perturbation in the curved spacetime, we can determine the effective potential  in a  similar way to the scalar field. Alternatively, the effective potential of spin one field can be obtained based on the formalism developed in Ref. \cite{Boonserm:2013dua},
\begin{eqnarray}
		V_{s=1}=f(r)\left[ \frac{\ell(\ell+1)}{h^2(r)}  \right] .
\end{eqnarray}

Now we summarize the above discussions. In the four-dimensional background with a wormhole-like metric,  Eqs.~\eqref{bbrn} and \eqref{shape}, the massless scalar ($s=0$) and electromagnetic ($s=1$) field perturbations can be described by the master equation Eq.~$\eqref{masterEq}$ with the spin-dependent effective potential,
\begin{eqnarray}
		V_{s}=f(r)\left\lbrace \frac{\ell(\ell+1)}{h^2(r)} +\frac{(1-s)}{h(r)}\frac{\d }{\d r}\left[ f(r)\frac{\d h(r)}{\d r}\right]  \right\rbrace.\label{RWpotential}
\end{eqnarray}

\section{Quasinormal spectra of the charged black-bounce}
\label{sec:QNM}

In the previous section, we derived the effective potential for the massless scalar and electromagnetic field perturbations in the four-dimensional background with a wormhole-like metric given by Eqs.~\eqref{bbrn} and \eqref{shape}.  Now we apply the shape function, see Eq.~\eqref{shape}, to the spin-dependent potential Eq.~\eqref{RWpotential} and compute the quasinormal spectrum in a charged black-bounce spacetime. 
There are several kinds of methods, such as the Leaver's method~\cite{Leaver:1985ax,Leaver:1990zz,Leaver:1992cf}, the time-domain integration method~\cite{Gundlach:1993tp},
and the semi-analytic WKB  method~\cite{Schutz:1985km,Iyer:1986np}, to compute the quasinormal mode frequencies. These methods have
their own advantages and disadvantages for different focuses. The Leaver's method is very accurate at a small
multipole number, but it has a disadvantage~\cite{Kokkotas:1999bd} compared to the WKB method, that is, the former cannot provide an intuition about the properties of quasinormal mode spectra. The direct integration method is more suitable for showing the time-domain profiles of quasinormal modes. The accuracy of the  WKB method may not be as good as that of the Leaver's method and the direct integration method when the multipole number is small, but the WKB method is more suitable for investigating quasinormal mode spectra as a whole. And the improved WKB method keeps a high degree of consistency with the Leaver's method in accuracy even for a small multiple number.

\subsection{Improved WKB method}


In order to overcome the disadvantage that the WKB method is less accurate than the Leaver's method at a small multiple number, we compute the quasinormal frequencies by using the improved WKB method~\cite{Matyjasek:2017psv}, i.e., the higher order WKB-Pad\'e approach in which the powerful technique of the Pad\'e transformation greatly improves the accuracy of computations. Incidentally, the WKB-Pad\'e approach has been widely used and its accuracy has been proved~\cite{Matyjasek:2019eeu,Konoplya:2003ii} to be consistent with the Leaver's method up to 24 decimal places at a small multiple number.
In order to obtain a complete and accurate spectrum, we first use the WKB-Pad\'e approach to extract the stable quasinormal mode frequencies
which are shown in Table \ref{table:s0} for the spin zero perturbation and Table \ref{table:s1} for the spin one perturbation in the unit of $m=1$. In the special case of $a=0$, the quasinormal frequencies for both spin zero and spin one perturbations recover the  Reissner-Nordstr\"om quasinormal frequencies. 
Our purpose is to explore the influence of bounce parameters on quasinormal mode spectra. We can see from the two tables that the spectra have the same behavior in large bounce parameters for different multiple
numbers, i.e., they become more stable because the absolute values of imaginary parts are decreasing. Moreover, we observe a significant decrease in both the real and imaginary parts of quasinormal frequencies when the bounce parameter is much bigger than the charge, $a\gg Q$. 

\begin{table}[h]
	\begin{ruledtabular}
		\begin{tabular}{ p{3em} ccccc}
			\multicolumn{6}{c}{$\ell=0$}\\
			\hline
			$a$ & $Q=0.1$ & $Q=0.3$ &  $Q=0.5$  & $Q=0.7$  & $0.9$  \\
			\hline
			$0$               & 0.110980 - 0.103232 $i$ & 0.110781 - 0.105527 $i$& 0.115632 - 0.105541 $i$ & 0.120872 - 0.105648 $i$  & 0.131885 - 0.100252 $i$ \\
			$0.1$             & 0.112182 - 0.103967 $i$ & 0.112576 - 0.103474 $i$& 0.114780 - 0.105693 $i$ & 0.121047 - 0.105530 $i$  & 0.131825 - 0.100105 $i$  \\
			$0.5$             & 0.110427 - 0.102556 $i$ & 0.112711 - 0.101803 $i$& 0.115609 - 0.102199 $i$ & 0.120628 - 0.102926 $i$  & 0.129820 - 0.097150 $i$  \\
			$1$               & 0.109065 - 0.098681 $i$ & 0.110501 - 0.097993 $i$& 0.113643 - 0.098061 $i$ & 0.121937 - 0.092844 $i$  & 0.120960 - 0.091151 $i$ \\
			$10$              & 0.060418 - 0.045365 $i$ & 0.060433 - 0.045425 $i$& 0.060463 - 0.045545 $i$ & 0.060511 - 0.045798 $i$  & 0.060730 - 0.046242 $i$ \\
			$50$              & 0.013350 - 0.011723 $i$ & 0.013350 - 0.011723 $i$& 0.013350 - 0.011725 $i$ & 0.013350 - 0.011727 $i$  & 0.013350 - 0.011729 $i$   \\
			$100$             & 0.006703 - 0.005939 $i$ & 0.006703 - 0.005939 $i$& 0.006703 - 0.005939 $i$ & 0.006703 - 0.005940 $i$  & 0.006704 - 0.005940  $i$   \\
			$500$             & 0.001333 - 0.001252 $i$ & 0.001333 - 0.001252 $i$& 0.001333 - 0.001252 $i$ & 0.001333 - 0.001252 $i$  & 0.001333 - 0.001252 $i$   \\
			\hline
			\multicolumn{6}{c}{$\ell=1$}\\
			\hline
			$0$               & 0.293435 - 0.097715 $i$ & 0.297554 - 0.098111 $i$& 0.306562 - 0.098803 $i$ & 0.322807 - 0.099362 $i$  & 0.352619 - 0.097170 $i$ \\
			$0.1$             & 0.293437 - 0.097660 $i$ & 0.297530 - 0.098054 $i$& 0.306549 - 0.098770 $i$ & 0.322786 - 0.099276 $i$  & 0.352604 - 0.097080 $i$  \\
			$0.5$             & 0.293363 - 0.096310 $i$ & 0.297455 - 0.096652 $i$& 0.306454 - 0.097211 $i$ & 0.322612 - 0.097510 $i$  & 0.352219 - 0.094878 $i$  \\
			$1$               & 0.293075 - 0.092018 $i$ & 0.297122 - 0.092166 $i$& 0.305994 - 0.092336 $i$ & 0.321860 - 0.091834 $i$  & 0.350633 - 0.087771 $i$ \\
			$10$              & 0.140575 - 0.039372 $i$ & 0.140634 - 0.039431 $i$& 0.140750 - 0.039545 $i$ & 0.140925 - 0.039723 $i$  & 0.141157 - 0.039960 $i$ \\
			$50$              & 0.030809 - 0.010045 $i$ & 0.030810 - 0.010045 $i$& 0.030810 - 0.010046 $i$ & 0.030812 - 0.010048 $i$  & 0.030813 - 0.010050 $i$   \\
			$100$             & 0.015568 - 0.005159 $i$ & 0.015568 - 0.005159 $i$& 0.015568 - 0.005159 $i$ & 0.015568 - 0.005159 $i$  & 0.015568 - 0.005160  $i$   \\
			$500$             & 0.003161 - 0.000971 $i$ & 0.003161 - 0.000971 $i$& 0.003161 - 0.000971 $i$ & 0.003161 - 0.000971 $i$  & 0.003161 - 0.000971 $i$   \\
			\hline
			\multicolumn{6}{c}{$\ell=2$}\\
			\hline
			$0$               & 0.484456 - 0.096812 $i$ & 0.491182 - 0.097219 $i$& 0.505967 - 0.097943 $i$ & 0.532563 - 0.098574 $i$  & 0.581959 - 0.096625 $i$ \\
			$0.1$             & 0.484455 - 0.096758 $i$ & 0.491180 - 0.097162 $i$& 0.505967 - 0.097880 $i$ & 0.532560 - 0.098495 $i$  & 0.581952 - 0.096534 $i$  \\
			$0.5$             & 0.484429 - 0.095445 $i$ & 0.491151 - 0.095793 $i$& 0.505919 - 0.096387 $i$ & 0.532469 - 0.096773 $i$  & 0.581741 - 0.094327 $i$  \\
			$1$               & 0.484285 - 0.091223 $i$ & 0.490976 - 0.091393 $i$& 0.505677 - 0.091575 $i$ & 0.532076 - 0.091155 $i$  & 0.580878 - 0.087099 $i$ \\
			$10$              & 0.227683 - 0.038247 $i$ & 0.227789 - 0.038306 $i$& 0.227999 - 0.038426 $i$ & 0.228315 - 0.038605 $i$  & 0.228735 - 0.038845 $i$ \\
			$50$              & 0.049895 - 0.009731 $i$ & 0.049896 - 0.009732 $i$& 0.049898 - 0.009733 $i$ & 0.049900 - 0.009734 $i$  & 0.049903 - 0.009736 $i$   \\
			$100$             & 0.025208 - 0.004996 $i$ & 0.025208 - 0.004996 $i$& 0.025209 - 0.004996 $i$ & 0.025209 - 0.004997 $i$  & 0.025209 - 0.004997 $i$   \\
			$500$             & 0.005088 - 0.000998 $i$ & 0.005088 - 0.000998 $i$& 0.005088 - 0.000998 $i$ & 0.005088 - 0.000998 $i$  & 0.005088 - 0.000998 $i$   \\
	\end{tabular}
\end{ruledtabular}
	\caption{The fundamental ($n=0$) quasinormal spectra of the spin zero field perturbation for the multipole number being from zero to two.  These frequencies are calculated by the WKB-Pad\'e  approach for various values of the bounce parameter and charge. The settings of $a$ and $Q$ are shown in the leftmost column and the top row, respectively. }\label{table:s0}
\end{table}

\begin{table}[h]
	\begin{ruledtabular}
		\begin{tabular}{ p{3em} ccccc}
			\multicolumn{6}{c}{$\ell=1$}\\
			\hline
			$a$ & $Q=0.1$ & $Q=0.3$ &  $Q=0.5$  & $Q=0.7$  & $0.9$  \\
			\hline
			$0$               & 0.248737 - 0.092554 $i$ & 0.252665 - 0.093036 $i$& 0.261386 - 0.093936 $i$ & 0.277399 - 0.094911 $i$  & 0.307983 - 0.093272 $i$ \\
			$0.1$             & 0.248775 - 0.092515 $i$ & 0.252680 - 0.092988 $i$& 0.261385 - 0.093910 $i$ & 0.277453 - 0.094885 $i$  & 0.308017 - 0.093179 $i$  \\
			$0.5$             & 0.249479 - 0.091299 $i$ & 0.253353 - 0.091773 $i$& 0.262127 - 0.092675 $i$ & 0.278217 - 0.093151 $i$  & 0.308812 - 0.091003 $i$  \\
			$1$               & 0.251589 - 0.087252 $i$ & 0.255592 - 0.087477 $i$& 0.264437 - 0.087769 $i$ & 0.280543 - 0.087525 $i$  & 0.310925 - 0.083491 $i$ \\
			$10$              & 0.117288 - 0.035018 $i$ & 0.117322 - 0.035059 $i$& 0.117391 - 0.035141 $i$ & 0.117493 - 0.035264 $i$  & 0.117629 - 0.035426 $i$ \\
			$50$              & 0.024984 - 0.008514 $i$ & 0.024984 - 0.008514 $i$& 0.024985 - 0.008515 $i$ & 0.024985 - 0.008516 $i$  & 0.024987 - 0.008517 $i$   \\
			$100$             & 0.012579 - 0.004349 $i$ & 0.012579 - 0.004349 $i$& 0.012579 - 0.004349 $i$ & 0.012579 - 0.004349 $i$  & 0.012579 - 0.004349  $i$   \\
			$500$             & 0.002536 - 0.000896 $i$ & 0.002536 - 0.000896 $i$& 0.002536 - 0.000896 $i$ & 0.002536 - 0.000896 $i$  & 0.002536 - 0.000896 $i$   \\
			\hline
			\multicolumn{6}{c}{$\ell=2$}\\
			\hline
			$0$               & 0.458393 - 0.095061 $i$ & 0.465012 - 0.095495 $i$& 0.479599 - 0.096281 $i$ & 0.505992 - 0.097021 $i$  & 0.555617 - 0.095216 $i$ \\
			$0.1$             & 0.458411 - 0.095007 $i$ & 0.465028 - 0.095439 $i$& 0.479616 - 0.096219 $i$ & 0.506003 - 0.096954 $i$  & 0.555636 - 0.095124 $i$  \\
			$0.5$             & 0.458779 - 0.093713 $i$ & 0.465405 - 0.094089 $i$& 0.480015 - 0.094740 $i$ & 0.506429 - 0.095219 $i$  & 0.556074 - 0.092898 $i$  \\
			$1$               & 0.459862 - 0.089513 $i$ & 0.466509 - 0.089702 $i$& 0.466509 - 0.089702 $i$ & 0.507616 - 0.089564 $i$  & 0.557215 - 0.085511 $i$ \\
			$10$              & 0.213525 - 0.036591 $i$ & 0.213613 - 0.036642 $i$& 0.213789 - 0.036745 $i$ & 0.214052 - 0.036899 $i$  & 0.214402 - 0.037104 $i$ \\
			$50$              & 0.046268 - 0.009113 $i$ & 0.046269 - 0.009113 $i$& 0.046270 - 0.009114 $i$ & 0.046272 - 0.009115 $i$  & 0.046275 - 0.009117 $i$   \\
			$100$             & 0.023343 - 0.004667 $i$ & 0.023343 - 0.004667 $i$& 0.023343 - 0.004667 $i$ & 0.023344 - 0.004667 $i$  & 0.023344 - 0.004668 $i$   \\
			$500$             & 0.004703 - 0.000952 $i$ & 0.004703 - 0.000952 $i$& 0.004703 - 0.000952 $i$ & 0.004703 - 0.000952 $i$  & 0.004703 - 0.000952 $i$   \\
	\end{tabular}
\end{ruledtabular}
	\caption{The fundamental ($n=0$) quasinormal spectra of the spin one field perturbation for the multipole number being one and two. These frequencies are calculated by the WKB-Pad\'e approach for various values of the bounce parameter and charge. The settings of $a$ and $Q$ are shown in the leftmost column and the top row, respectively. }\label{table:s1}
\end{table}
\subsection{Leaver's method}

Before we apply the Leaver's method to make numerical calculations of quasinormal spectra, we have to do the following analytical analyses because the effective potential, Eq.~(\ref{RWpotential}), contains the terms with fractional powers. 
In order to overcome the problem, we make a coordinate transformation, $r\to h$, where $h\in[a,\infty)$.
By using the newly defined symbols, 
\begin{equation}
	F\coloneqq \frac{f}{u} =\frac{\sqrt{h^2-a^2} }{h^3}\left(h^2-2h+Q^2\right),\qquad u\coloneqq \frac{\d r}{\d h}=\frac{h}{\sqrt{h^2-a^2}},	
\end{equation}
we rewrite the master wave equation, Eq.~(\ref{masterEq}), to be
\begin{equation}
	\label{eq:master-step}
	-F(h) \frac{\d}{\d h}\left[
	F(h) \frac{\d}{\d h}
	\Psi_s(h)\right]+V_s(h) \Psi_s(h) =\omega^2 \Psi_s(h),
\end{equation}
where the corresponding boundary conditions become
\begin{subequations}
	\label{eq:boundary}
	\begin{equation}
		\Psi_s\sim e^{i \omega h_*}\to e^{i h \omega } h^{2 i \omega },\quad
		h\to \infty,
	\end{equation}
	\begin{equation}
	\label{eq:boundary-horizon}
		\Psi_s\sim e^{-i \omega h_*}\to (h-h_+)^{\beta},\quad
		h\to h_+,
	\end{equation}
\end{subequations}
with the corresponding ``tortoise'' coordinate defined by 
\begin{equation}
h_*\coloneqq \int \frac{\d h}{F(h)},
\end{equation}
the corresponding horizons located at
\begin{equation} 
h_\pm=1\pm \sqrt{1-Q^2},\label{hphmq} 
\end{equation}
and the parameter $\beta$ expressed by
\begin{equation}
\label{eq:root}
\beta=\frac{-i\omega h_+^3  }{2\left( h_+-1\right) \sqrt{h_+^2-a^2}}.
\end{equation}
Further, we construct a series solution in terms of the rational function, $z\coloneqq \frac{h-h_+}{h-h_-}$,
\begin{equation}
	\label{eq:serise}
	\Psi_s=A(h,\omega) \sum_{n=0}^\infty c_n z^{n},
\end{equation}
where the prefactor 
\begin{equation}
	A(h,\omega)=e^{i h \omega } h^{2 i \omega } z^{\beta}
\end{equation}
is determined by the boundary conditions, Eq.\ \eqref{eq:boundary}. 

Now we can find the recursion relations by substituting Eq.\ \eqref{eq:serise} into Eq.\ \eqref{eq:master-step}.
The result is a seven-term recursion relation, as detailed in Appendix~\ref{app:recurrence}.
Next, we perform the Leaver's algorithms \cite{Leaver:1990zz,Onozawa:1995vu,Jing:2005cb,Jing:2008an,Jing:2005dt} for accurately calculating the QNMs, 
i.e., we extract the quasinormal spectra of the charged black-bounce by following the two ways.
\begin{enumerate}
    \item {\bf The Hill's determinant method}. Here we set the order of determinant to be eleven, $n= 11$. This method is more effective than the continued fraction approach because the later becomes inconceivably complicated as the order of the continued fraction increases. For the details, see Appendix~\ref{app:recurrence}.
    \item {\bf The continued fraction approach}. We reduce the seven-term recurrence relation to a three-term one by the quadruple Gaussian elimination, see Appendix~\ref{app:gaussian}. The three-term recurrence relation can further be used to calculate the quasinormal spectra by the continued fraction approach.
\end{enumerate}

Table~\ref{leaver:s0} and Table~\ref{leaver:s1} show the quasinormal spectra of spin zero and spin one field perturbations computed by the Hill's determinant method, respectively. 
They are also checked by the continued fraction approach.
In addition, we compare them with those computed by the improved WKB method. 
For the two types of spin fields, the data from the improved WKB method and Leaver's method show a high-degree agreement when the bounce parameter is getting smaller. 

The Leaver's method also shows the high accuracy at a small multipole number, but it fails to give an accurate and stable frequency for a large bounce parameter, $a$, in contrast to the improved WKB approach. 
This limitation of the Leaver's method is most likely caused by the black-bounce model itself in which 
the model has transformed from a regular black hole to a wormhole with the vanishing horizon when the bounce parameter becomes large.
In other words, the boundary condition in Eq.\ \eqref{eq:boundary-horizon} is no longer valid when no horizons exist, that is, the effective potential $V_s(h)$ has no intersection with the $h$-axis. 
Subsequently, the calculation goes beyond its validity as the bounce parameter crosses the critical value between the black holes and wormholes.

\begin{table}[h]
	\begin{ruledtabular}
		\begin{tabular}{ p{3em} ccccc}
			\multicolumn{6}{c}{$\ell=0$}\\
			\hline
			$a$ & $Q=0.1$ & $Q=0.3$ &  $Q=0.5$  & $Q=0.7$  & $0.9$  \\
			\hline
			$0$    & 0.110620 - 0.106498 $i$  & 0.111981 - 0.106878 $i$& 0.114866 - 0.107375 $i$ & 0.119820 - 0.106704 $i$  & 0.131114 - 0.098885 $i$ \\
			$0.1$  & 0.110605 - 0.106426 $i$ & 0.111965 - 0.106803 $i$& 0.114846 - 0.107292 $i$ & 0.119792 - 0.106608 $i$  & 0.131060 - 0.098780 $i$  \\
			$0.5$  & 0.110230 - 0.104700 $i$ & 0.111548 - 0.104998 $i$& 0.114330 - 0.105306 $i$ & 0.119080 - 0.104303 $i$  & 0.129665 - 0.096306 $i$  \\
			$1$     & 0.108277 - 0.099859 $i$ & 0.109343 - 0.099928 $i$& 0.111500 - 0.099687 $i$ & 0.114925 - 0.097549 $i$  & 0.122514 - 0.088390 $i$ \\
			$10$   & 0.002252 - 0.048342 $i$ & 0.002250 - 0.050383 $i$& 0.002262 - 0.055250 $i$ & 0.002353 - 0.065750 $i$  & 0.002953 - 0.097677 $i$ \\
			$50$   & 0.000292 - 0.054529 $i$ & 0.000300 - 0.056482 $i$& 0.000319 - 0.061174 $i$ & 0.000363 - 0.071430 $i$  & 0.000522 - 0.103263 $i$   \\
			$100$  & 0.000144 - 0.054713 $i$ & 0.000148 - 0.056664 $i$& 0.000157 - 0.061353 $i$ & 0.000180 - 0.071604 $i$  & 0.000260 - 0.103438  $i$   \\
			$500$  & 0.000029 - 0.054772 $i$ & 0.000030 - 0.056723 $i$& 0.000031 - 0.061410 $i$ & 0.000036 - 0.071659 $i$  & 0.000052 - 0.103494 $i$   \\
			\hline
			\multicolumn{6}{c}{$\ell=1$}\\
			\hline
			$0$    & 0.293390 - 0.097716 $i$ & 0.297489 - 0.098105 $i$& 0.306503 - 0.098777 $i$ & 0.322709 - 0.099261 $i$  & 0.352881 - 0.096925 $i$ \\
			$0.1$  & 0.293388 - 0.097660 $i$ & 0.297487 - 0.098047 $i$& 0.306499 - 0.098713 $i$ & 0.322702 - 0.099188 $i$  & 0.352865 - 0.096834 $i$  \\
			$0.5$  & 0.293330- 0.096310 $i$ & 0.297417 - 0.096641 $i$& 0.306397 - 0.097184 $i$ & 0.322525 - 0.097425 $i$  & 0.352461 - 0.094633 $i$  \\
			$1$     & 0.292937 - 0.092070 $i$ & 0.296955 - 0.092219 $i$& 0.305761 - 0.092336 $i$ & 0.321489 - 0.091690 $i$  & 0.351138 - 0.086528 $i$ \\
			$10$   & 0.019957 - 0.255472 $i$ & 0.019941 - 0.261539 $i$& 0.020063 - 0.276306 $i$ & 0.020877 - 0.308994 $i$  & 0.002310 - 0.039827 $i$ \\
			$50$  & 0.000296 - 0.050628 $i$ & 0.000304 - 0.052710 $i$& 0.000323 - 0.057679 $i$ & 0.000368 - 0.068412 $i$  & 0.000529 - 0.101121 $i$   \\
			$100$ & 0.000144 - 0.053744 $i$ & 0.000148 - 0.055727 $i$& 0.000158 - 0.060483 $i$ & 0.000180 - 0.070852 $i$  & 0.000261 - 0.102904  $i$   \\
			$500$ & 0.000029 - 0.054734 $i$ & 0.000029 - 0.056685 $i$& 0.000031 - 0.061375 $i$ & 0.000036 - 0.071629 $i$  & 0.000052 - 0.103473 $i$   \\
			\hline
			\multicolumn{6}{c}{$\ell=2$}\\
			\hline
			$0$     & 0.484454 - 0.096806 $i$ & 0.491178 - 0.097212 $i$& 0.505967 - 0.097931 $i$ & 0.532574 - 0.098547 $i$  & 0.582100 - 0.096679 $i$ \\
			$0.1$    & 0.484453 - 0.096751 $i$ & 0.491177 - 0.097155 $i$& 0.505965 - 0.097869 $i$ & 0.532571 - 0.098475 $i$  & 0.582092 - 0.096588 $i$  \\
			$0.5$   & 0.484428 - 0.095438 $i$ & 0.491145 - 0.095787 $i$& 0.505916 - 0.096376 $i$ & 0.532480 - 0.096743 $i$  & 0.581888 - 0.094371 $i$  \\
			$1$      & 0.484239 - 0.091237 $i$ & 0.490921 - 0.091402 $i$& 0.505603 - 0.091561 $i$ & 0.531980 - 0.091043 $i$  & 0.581740 - 0.086938 $i$ \\
			$10$    & 0.020684 - 0.192118 $i$ & 0.020948 - 0.200169 $i$& 0.021635 - 0.219038 $i$ & 0.023363 - 0.258366 $i$  & 0.030441 - 0.369276 $i$ \\
			$50$   & 0.000293 - 0.042766 $i$ & 0.000303 - 0.045111 $i$& 0.000325 - 0.050640 $i$ & 0.000374 - 0.062339 $i$  & 0.000540 - 0.096815 $i$   \\
			$100$  & 0.000145 - 0.051803 $i$ & 0.000149 - 0.053849 $i$& 0.000159 - 0.058742 $i$ & 0.000182 - 0.069347 $i$  & 0.000262 - 0.101833 $i$   \\
			$500$  & 0.000029 - 0.054656 $i$ & 0.000029 - 0.056611 $i$& 0.000031 - 0.061306 $i$ & 0.000036 - 0.071569 $i$  & 0.000052 - 0.103430 $i$   \\
		\end{tabular}
	\end{ruledtabular}
	\caption{The fundamental ($n=0$) quasinormal spectra of the spin zero field perturbation for the multipole number being from zero to two.  These frequencies are calculated by the Leaver's method for various values of the bounce parameter and charge. The settings of $a$ and $Q$ are shown in the leftmost column and the top row, respectively. }\label{leaver:s0}
\end{table}
\begin{table}[h]
	\begin{ruledtabular}
		\begin{tabular}{ p{3em} ccccc}
			\multicolumn{6}{c}{$\ell=1$}\\
			\hline
			$a$ & $Q=0.1$ & $Q=0.3$ &  $Q=0.5$  &  $Q=0.7$  & $0.9$   \\
			\hline
			$0$               & 0.248678 - 0.092485 $i$ & 0.252617 - 0.092957 $i$& 0.261356 - 0.093822 $i$ & 0.277384 - 0.094686 $i$  & 0.308523 - 0.093206 $i$  \\
			$0.1$             & 0.248708 - 0.092436 $i$ & 0.252647 - 0.092905 $i$& 0.261388 - 0.093765 $i$ & 0.277418 - 0.094617 $i$  & 0.308558 - 0.093115 $i$   \\
			$0.5$             & 0.249417 - 0.091227 $i$ & 0.253373 - 0.091639 $i$& 0.262149 - 0.092364 $i$ & 0.278224 - 0.092952 $i$  & 0.309360 - 0.090890 $i$    \\
			$1$               & 0.251321 - 0.087249 $i$ & 0.255298 - 0.087445 $i$& 0.264102 - 0.087652 $i$ & 0.280187 - 0.087125 $i$  & 0.312038 - 0.082277 $i$  \\
			$10$              & 0.039266 - 0.158839 $i$ & 0.039335 - 0.162396 $i$& 0.039626 - 0.170899 $i$ & 0.040654 - 0.189154 $i$  & 0.045816 - 0.242896 $i$ \\
			$50$              &  ***  &  ***&  ***  & *** & ***  \\

			\hline
			\multicolumn{6}{c}{$\ell=2$}\\
			\hline
			$0$               & 0.458394 - 0.095051 $i$ & 0.465012 - 0.095484 $i$& 0.479606 - 0.096265 $i$ & 0.506011 - 0.096997 $i$  & 0.555776 - 0.095340 $i$  \\
			$0.1$             & 0.458410 - 0.094998 $i$ & 0.465028 - 0.095429 $i$& 0.479622 - 0.096204 $i$  & 0.506029 - 0.096926 $i$  & 0.555796 - 0.095248 $i$  \\
			$0.5$             & 0.458779 - 0.093704 $i$ & 0.465406 - 0.094078 $i$& 0.480018 - 0.094726 $i$ & 0.506451 - 0.095199 $i$  & 0.556245 - 0.093006 $i$  \\
			$1$               & 0.459803 - 0.089521 $i$ & 0.466442 - 0.089701 $i$& 0.481073 - 0.089893 $i$  & 0.507538 - 0.089419 $i$  & 0.558181 - 0.085457 $i$\\
			$10$              & 0.022139 - 0.106579 $i$ & 0.022917 - 0.112669 $i$& 0.024702 - 0.126604 $i$  & 0.028331 - 0.154274 $i$  & 0.038625 - 0.224482 $i$\\
			$50$              & 0.006374 - 0.137041 $i$ & 0.006432 - 0.141000 $i$& 0.006586 - 0.150454 $i$ & *** & ***   \\
			$100$             & ***  &  *** &  *** & *** & ***  \\
		
		\end{tabular}
	\end{ruledtabular}
	\caption{The fundamental ($n=0$) quasinormal spectra of the spin one field perturbation for the multipole number being one and two. These frequencies are calculated by the Leaver's method for various values of the bounce parameter and charge. The settings of $a$ and $Q$ are shown in the leftmost column and the top row, respectively. }\label{leaver:s1}
\end{table}

\section{Gray-body factors}
\label{sec:GBF}
\begin{figure}[htbp]
	\centering
	\includegraphics[width=0.5\linewidth]{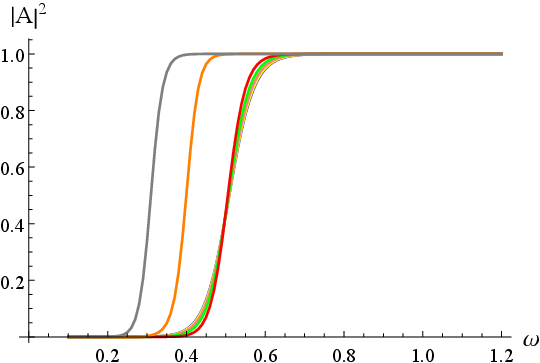}\includegraphics[width=0.5\linewidth]{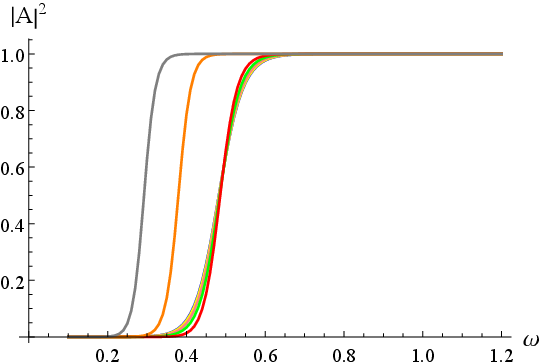}
	\caption{Gray-body factors of spin zero (left) and spin one (right) fields with $\ell=2$, $m=1$, and $Q=0.5$. The bounce parameter takes seven different values: $a=0$ (blue), $a=0.5$ (yellow), $a=1$ (pink), $a=1.5$ (green), $a=2$ (red), $a=5$ (orange), and $a=7$ (gray).}
	\label{gbfactor}
\end{figure}

In order to investigate the bounce corrections to the transmission probability of particles scattered by the effective potential, we should analyze the gray-body factors of perturbation fields. We need to solve the wave equation Eq.~(\ref{masterEq}) with the scattering boundary conditions,
\begin{eqnarray}
		\Psi&=&{\cal T}e^{-i\omega r_*}, \qquad \quad \quad\,\,   r_* \rightarrow - \infty,  \nonumber\\
    	\Psi&=&e^{-i\omega r_*} + {\cal R} e^{i\omega r_*}, \quad   r_* \rightarrow + \infty,
\end{eqnarray}
where ${\cal T}$ and ${\cal R}$ are the transmission and reflection coefficients, respectively. The boundary conditions allow the incoming wave from infinity. For a given multipole number $\ell$, one has~\cite{Iyer:1986np}
\begin{eqnarray}
	|A_{\ell}|^2=1-|{\cal R}_{\ell}|^2=|{\cal T}_{\ell}|^2,\quad \quad |{\cal R}_{\ell}|^2=\left(1+e^{-2i\pi {\mathcal K}}\right)^{-1},
\end{eqnarray} 
where $\mathcal K$ is determined by 
\begin{eqnarray}
	{\mathcal K}=i\frac{\omega^2-V_0}{\sqrt{-2V_0''}}-\sum_{i=2}^{i=6}\Lambda_i({\mathcal K}),
\end{eqnarray}
$V_0$ is the maximum of the potential, $V_0''$ is the second derivative with respect to the tortoise coordinate at the location where the potential takes its maximum, and $\Lambda_i$'s denote \cite{Schutz:1985km,Konoplya:2003ii,Konoplya:2019hlu} the higher WKB corrections. Fig. \ref{gbfactor} shows the fact that a particle with a larger frequency (larger energy) is more likely to pass through the potential barrier, i.e. it has a higher gray-body factor. Additionally, a large bounce parameter $a$ also leads to a higher gray-body factor, which can be seen from Fig. \ref{gbfactor}. That is, the contour of gray-body factors moves left when $a$ increases from zero. It is worth noting that it seems somewhat difficult to distinguish the contours when $a<2$, which shows that the gray-body factors are almost  independent of a small bounce parameter.
However, when $a$ increases, especially up to $a>m\pm\sqrt{m^2-Q^2}$ (the geometry is a traversable wormhole), we can clearly observe the increased gray-body factors. In this sense, we can conclude that the corrections of bounce parameters to the Reissner-Nordstr\"om black hole lead to  increasing in gray-body factors.

\section{Conclusions}
	\label{sec:con}
In this work we have investigated the weak gravitational lensing, quasinormal spectra, and gray-body factors of the Reissner-Nordstr\"om spacetime corrected by bounce parameters. By applying the Gauss-Bonnet theorem to the optical geometry, we find that there exists a bounce correction of the Morse potential to suppress the divergence of the deflection angle in the Reissner-Nordstr\"om spacetime.  Moreover, we derive the master wave equations with the spin-dependent potentials for massless scalar and electromagnetic field perturbations. We then observe that the spectra have the same behavior in large bounce parameters for different multipole numbers and they decrease significantly if $a\gg Q$. Furthermore, we compare the results computed by the Leaver's method with those by the improved WKB method, and find that the quasinormal frequencies show a high-degree agreement in the two methods at a smaller bounce parameter for the scalar and vector field perturbations. Finally, the results of scattering problems suggest that the corrections of bounce parameters introduced into the Reissner-Nordstr\"om spacetime lead to increasing in the gray-body factors of perturbation fields.

\section*{Acknowledgments}
The authors would like to thank 
Qiyuan Pan for sharing his Mathematica codes of the Leaver's method and
the anonymous referee for the helpful comments that
improve this work greatly.
This work was supported in part by the National Natural Science Foundation of China under Grant  Nos. 11675081 and 12175108.

\appendix

\section{ Recurrence relations}
\label{app:recurrence}

In order to find the recurrence relations, we substitute Eq.\ \eqref{eq:serise} into Eq.\ \eqref{eq:master-step}
and obtain the characteristic equation at first,
\begin{equation}
\label{eq:ch-eq}
	z^0:\quad 
	4 {\beta}^2 \left(h_+-1\right){}^2 \left(h_+^2-a^2\right) +h_+^6 \omega ^2=0,
\end{equation}
which is also known as the {\em indicial polynomial} 
and will be used to determine the parameter $\beta$.
The root $\beta$ from Eq.\ \eqref{eq:ch-eq} must be consistent with the one from the boundary condition, Eq.\ \eqref{eq:root}.

Then, we derive the other equations order by order,
\begin{equation}
\label{eq:rec2}
	z^1:\quad
	\beta_0 c_0+\alpha_0 c_1=0;
\end{equation}
\begin{equation}
\label{eq:rec3}
	z^2:\quad
	\gamma_0 c_0+\beta_1 c_1+\alpha_1 c_2=0;
\end{equation}
\begin{equation}
\label{eq:rec4}
	z^3:\quad
	\delta_0 c_0+\gamma_1 c_1+\beta_2 c_2+\alpha_2 c_3=0;
\end{equation}
\begin{equation}
\label{eq:rec5}
	z^4:\quad
	\zeta_0 c_0+\delta_1 c_1+\gamma_2 c_2+\beta_3 c_3+\alpha_3 c_4=0;
\end{equation}
\begin{equation}
\label{eq:rec6}
	z^5:\quad
	\eta_0 c_0+\zeta_1 c_1+\delta_2 c_2+\gamma_3 c_3+\beta_4 c_4+\alpha_4 c_5=0;
\end{equation}
\begin{equation}
\label{eq:rec7}
	z^n:\quad
	\kappa_{n-6} c_{n-6}+\eta_{n-5} c_{n-5}+\zeta_{n-4} c_{n-4}+\delta_{n-3} c_{n-3}+\gamma_{n-2} c_{n-2}+\beta_{n-1} c_{n-1}+\alpha_{n-1} c_n=0,\quad n\ge 6.
\end{equation}
The last one is a seven-term recurrence relation. The coefficients in the above recurrence relations are given as follows,\footnote{Note that our subscript labels of the coefficients are slightly different from those used in Ref.\ \cite{Leaver:1992cf}.
In our notation all the subscripts start from zero, $n=0, 1, 2, \dots$.}
\begin{equation}
	\alpha_n=4 \left(h_+-1\right)^2 h_+^2 \left(a^2-h_+^2\right) (b+n+1)^2-h_+^8 \omega ^2,
\end{equation}
\begin{equation}
	\begin{split}
		\beta_n=
		&-2 \Big\{
		4 a^2 \left(h_+-1\right)^2 h_+ 
		\Big[b^2 h_++2 b^2-i 
		\left(h_+-1\right) 
		\left(h_++2\right) 
		\omega  (2 b+2 n+1)\\
		&		+2 b h_+ n+3 b h_++4 b n-3 b+h_+ n^2+3 h_+ n+h_+ 
		\epsilon +2 n^2-3 n-\epsilon 
		\Big]\\
		&	+h_+^3 
		\Big[
		4 i \left(h_+-1\right)^3 
		\left(h_++2\right) \omega  (2 b+2 n+1)
		+\left(5 h_+-8\right) h_+^4 \omega ^2\\
		&		+2 
		\left(h_+-1\right)^2
		\left[2 (b+n) 
		\left(b 
		\left(h_+-4\right)+h_+ (n-2)-4 n+2
		\right)-h_+ \lambda -2 
		\left(h_+-1\right) \epsilon 
		\right]
		\Big]
		\Big\},
	\end{split}
\end{equation}
\begin{equation}
	\begin{split}
		\gamma_n=
		&h_+^2 
		\Big\{
		4 \left(h_+-1\right)^2 
		\Big[b^2 
		\left(h_+ 
		\left(h_++8\right)-24
		\right)+2 b 
		\left(h_+ \left(h_++8\right)-24\right) n\\
		&+8 b \left(h_+-3\right) \left(h_+-1\right)
		+4 \left(h_+-2\right) h_+ \lambda 
		+\left(h_+ \left(h_++8\right)-24\right) n^2
		+8 \left(h_+-3\right) \left(h_+-1\right) n
		-4 \left(h_+-1\right) \epsilon 
		\Big]\\
		&-16 i 
		\left(h_+-1\right)^3 \omega  
		\left[4 b 
		\left(
		\left(h_+-1\right) h_+-3
		\right)+4 
		\left(\left(h_+-1\right) h_+-3\right) n
		+\left(h_+-4\right) h_+
		\right]\\
		&+
		\left[
		64-\left(h_+-2\right) h_+ 
		\left(h_+ 
		\left(h_+ 
		\left(h_+ \left(31 h_+-82\right)+44\right)+24
		\right)-96
		\right)
		\right] \omega ^2
		\Big\}\\
		&-4 a^2 \left(h_+-1\right)^2 
		\Big\{2 b 
		\left[
		h_+^2 n-12 h_+ n+8 i \left(h_+-1\right) \left(2 h_++1\right) \omega -6 h_+^2-4 n+6
		\right]\\
		&+b^2 
		\left[\left(h_+-12\right) h_+-4\right]
		+\left[\left(h_+-12\right) h_+-4\right] n^2+4 i 
		\left(h_+-1\right) n \left(8 h_+ \omega +3 i h_++4 \omega +3 i\right)\\
		&+4 \left(h_+-1\right) 
		\left[
		\left(h_+-1\right) 
		\left(h_++2\right)^2 \omega ^2
		+i \left(h_+ \left(2 h_++7\right)-6\right) \omega +\left(2-3 h_+\right) \epsilon 
		\right]
		\Big\},
	\end{split}
\end{equation}
\begin{equation}
	\begin{split}
		\delta_n=&4 \Big\{
		4 a^2 \left(h_+-1\right)^2 
		\Big[b \left(2 h_+^2 n-4 h_+ n-2 i 
		\left(h_+-4\right) 
		\left(h_+-1\right) 
		\left(h_++2\right) \omega -9 h_+-8 n+9
		\right)\\
		&+\left(\left(h_+-2\right) h_+-4\right) n^2+
		\left(h_+-1\right) n 
		\left(-9-2 i \left(h_+-4\right) \left(h_++2\right) \omega \right)\\
		&b^2 
		\left(\left(h_+-2\right) h_+-4\right)
		-\left(h_+-1\right)^2 
		\left(
		5 \epsilon +2 \omega  
		\left(\left(h_+-4\right) \left(h_++2\right) \omega -9 i\right)
		\right)
		\Big]\\
		&+\left(h_+-2\right) h_+ 
		\Big[
		2 \left(h_+-1\right)^2 
		\Big(
		2 b^2 \left(\left(h_+-2\right) h_+-4\right)
		+4 b \left(h_+ \left(\left(h_+-2\right) n-3\right)-4 n+3\right)\\
		&+3 \left(h_+-2\right) h_+ \lambda 
		+2 n \left(h_+ \left(\left(h_+-2\right) n-6\right)-4 n+6\right)
		-2 \left(h_+-1\right)^2 \epsilon 
		\Big)\\
		&-24 i \left(h_+-1\right)^3 \omega  
		\left(
		b \left(\left(h_+-2\right) h_+-2\right)
		+\left(\left(h_+-2\right) h_+-2\right) n-h_++1
		\right)\\
		&+
		\left(
		32-\left(h_+-2\right) h_+ 
		\left(
		\left(h_+-2\right) h_+ 
		\left(11 \left(h_+-2\right) h_++8\right)-40
		\right)
		\right) \omega ^2
		\Big]
		\Big\},
	\end{split}
\end{equation}
\begin{equation}
	\begin{split}
		\zeta_n=
		&4 b^2 \left(h_+-1\right)^2 
		\left\{
		\left(h_+-2\right)^2 
		\left[\left(h_+-12\right) h_+-4\right]
		-a^2 \left[h_+ \left(h_++8\right)-24\right]
		\right\}\\
		&+8 b \left(h_+-1\right)^2 
		\Big\{
		a^2 \left[-h_+^2 (n-16 i \omega +6)-8 h_+ (n+7 i \omega -3)+24 n+40 i \omega -18\right]\\
		&+\left(h_+-2\right)^2 
		\left[h_+^2 n-12 h_+ n-8 i 
		\left(h_+ 
		\left(\left(h_+-4\right) h_++2
		\right)+1
		\right) \omega -4 h_+^2-4 n+4
		\right]
		\Big\}\\
		&+4 \left(h_+-1\right)^2 n^2 
		\left\{
		\left(h_+-2\right)^2 
		\left[
		\left(h_+-12\right) h_+-4
		\right]
		-a^2 \left[h_+ \left(h_++8\right)-24\right]
		\right\}\\
		&+16 \left(h_+-1\right)^3 n 
		\left\{
		a^2 \left[h_+ (-3+8 i \omega )-20 i \omega +9\right)]
		+2 \left(h_+-2\right)^2 
		\left[
		-2 i \left(\left(h_+-3\right) h_+-1\right) 
		\omega -h_+-1
		\right]
		\right\}\\
		&-\omega ^2 
		\left(h_+-2\right)^2 \left\{
		16 a^2 \left(h_++2\right)^2 
		\left(h_+-1\right)^4
		+
		\left[
		\left(h_+-2\right) h_+ 
		\left(
		h_+ 
		\left(
		h_+ 
		\left(
		h_+ 
		\left(31 h_+-166\right)+296
		\right)-208
		\right)-32
		\right)-64
		\right]
		\right\}\\
		&+16 i \left(h_+-1\right)^3 \omega  
		\left\{a^2 
		\left[h_+ \left(2 h_+-15\right)+16\right]
		+\left(h_++2\right) \left(h_+-2\right)^3
		\right\}\\
		&+16 \left(h_+-1\right)^2 
		\left\{
		\left(h_+-1\right) \epsilon  
		\left[a^2 
		\left(3 h_+-4\right)+\left(h_+-2\right)^2
		\right]+h_+ \left(h_+-2\right)^3 \lambda 
		\right\},
	\end{split}
\end{equation}
\begin{equation}
	\begin{split}
		\eta_n=
		&-2 \left(h_+-2\right) 
		\Big\{
		4 b^2 \left(h_+-1\right)^2 
		\left[a^2 \left(h_+-4\right)
		+\left(h_+-2\right)^2 
		\left(h_++2\right)
		\right]\\
		&+4 b \left(h_+-1\right)^2 
		\Big[
		a^2 \left(
		2 h_+ n-2 i \left(h_+-4\right) 
		\left(h_+-1\right) \omega -3 h_+-8 n+3
		\right)\\
		&+2 \left(h_+-2\right)^2 
		\left(h_+ n+i 
		\left(h_+-4\right) 
		\left(h_+-1\right) \omega +h_++2 n-1
		\right)
		\Big]\\
		&+4 \left(h_+-1\right)^2 n^2 
		\left[a^2 
		\left(h_+-4\right)
		+\left(h_+-2\right)^2
		\left(h_++2\right)
		\right]\\
		&+4 \left(h_+-1\right)^3 n 
		\left[a^2 \left(-3-2 i \left(h_+-4\right) \omega \right)
		+2 \left(h_+-2\right)^2 
		\left(1+i \left(h_+-4\right) \omega \right)\right]\\
		&-4 i 
		\left(h_+-4\right) 
		\left(h_+-1\right)^3 \omega  
		\left[-a^2+(h_+-2)^2\right] 
		-4 \left(h_+-1\right)^3 \epsilon  
		\left[-a^2+(h_+-2)^2\right]\\
		&-2 \left(h_+-1\right)^2 
		\left(h_+-2\right)^3 \lambda
		+\left(5 h_+-2\right) 
		\left(h_+-2\right)^6 \omega ^2	
		\Big\},
	\end{split}
\end{equation}
and
\begin{equation}
	\begin{split}
		\kappa_n=
		4 \left(h_+-2\right)^2 
		\left(h_+-1\right)^2 
		\left[a^2-(h_+-2)^2\right]
		(b+n)^2
		-\left(h_+-2\right)^8 \omega ^2
	\end{split},
\end{equation}
where we have set $m=1$,
used the relations, $\lambda= \ell(\ell+1)$ and
$\epsilon=1-s$,
and replaced $Q$ and $h_-$ by $h_+$ with the help of Eq.~(\ref{hphmq}) in order to simplify the notations.

The Leaver's matrix of order $n$ then reads
\begin{equation}
	\mathcal{M}(n)=\begin{pmatrix}
		\beta_0 & \alpha_0 &  &  &  &  &  &  &   \\
		\gamma_0 & \beta_1 & \alpha_1 &  &  &  &  &  &   \\
		\delta_0& \gamma_1 & \beta_2 & \alpha_2 &  &  &  &  &   \\
		\zeta_0 & \delta_1 & \gamma_2 & \beta_3 & \alpha_3 &  &  &  &   \\
		\eta_0 & \zeta_1 & \delta_2 & \gamma_3 & \beta_4 & \alpha_4 &  &  &   \\
		\kappa_0 & \eta_1 & \zeta_2 & \delta_3 & \gamma_4& \beta_5 & \alpha_5 &  &   \\
		& \ddots & \ddots &\ddots & \ddots & \ddots & \ddots & \ddots &   \\
		&  & \kappa_{n-6} & \eta_{n-5} & \zeta_{n-4} & \delta_{n-3} & \gamma_{n-2} & \beta_{n-1} & \alpha_{n-1}  \\
		&  &  & \kappa_{n-5} & \eta_{n-4} & \zeta_{n-3} & \delta_{n-2} & \gamma_{n-1} & \beta_{n}  \\
	\end{pmatrix},
\end{equation}
and the quasinormal spectrum can be computed by the vanishing Hill's determinant, i.e.\
\begin{equation}
\label{eq:determi}
		\det \mathcal{M}(n)=0.
\end{equation}

\section{Gaussian elimination}
\label{app:gaussian}

To utilize the continued fraction approach, we have to reduce the recurrence relations that have more than three terms in Eqs.\ (\ref{eq:rec2})-(\ref{eq:rec7}) to three-term ones. 
This can be realized in terms of the Gaussian elimination (GE) \cite{Leaver:1992cf,Onozawa:1995vu}, where the  quadruple GE, at the most, will be adopted for Eq.\ (\ref{eq:rec7}).

The single GE provides a six-term recurrence relation, i.e., $7\to6$.
The coefficient $\kappa_n$ is eliminated
and Eq.\ (\ref{eq:rec7}) is reduced to be
\begin{equation}
	\eta'_{n-5} c_{n-5}+\zeta'_{n-4} c_{n-4}+\delta'_{n-3} c_{n-3}+\gamma'_{n-2} c_{n-2}+\beta'_{n-1} c_{n-1}+\alpha'_{n-1} c_n=0,\label{sevtosix}
\end{equation}
where the new coefficients are given by
\begin{equation}
	\begin{split}
		\eta'_0 & =\eta_0;\\
		\beta'_n & =\beta_n,\quad n<5;\\
		\gamma'_n & =\gamma_n,\quad n<4;\\
		\delta'_n & =\delta_n,\quad n<3;\\
		\zeta'_n & =\zeta_n,\quad n<2;\\
	\end{split}
\end{equation}
and 
\begin{subequations}
	\begin{equation}
		\alpha'_n=\alpha_n;
	\end{equation}
	\begin{equation}
		\beta'_n=\beta_n-\frac{\kappa_{n-5}}{\eta'_{n-5}}\alpha'_{n-1},\qquad
		n\ge 5;
	\end{equation}
	\begin{equation}
		\gamma'_n=\gamma_n-\frac{\kappa_{n-4}}{\eta'_{n-4}}\beta'_n,\qquad
		n\ge 4;
	\end{equation}
	\begin{equation}
		\delta'_n=\delta_n-\frac{\kappa_{n-3}}{\eta'_{n-3}}\gamma'_n,\qquad
		n\ge 3;
	\end{equation}
	\begin{equation}
		\zeta'_n=\zeta_n-\frac{\kappa_{n-2}}{\eta'_{n-2}}\delta'_n,\qquad
		n\ge 2;
	\end{equation}
	\begin{equation}
		\eta'_n=\eta_n-\frac{\kappa_{n-1}}{\eta'_{n-1}}\zeta'_n
		,\qquad
		n\ge 1.
	\end{equation}
\end{subequations}

Similarly, the double GE provides
$6\to5$. 
The coefficient $\eta'_n$ is eliminated in the recurrence relation, Eq.~(\ref{sevtosix}),
\begin{equation}
	\zeta''_{n-4} c_{n-4}+\delta''_{n-3} c_{n-3}+\gamma''_{n-2} c_{n-2}+\beta''_{n-1} c_{n-1}+\alpha''_{n-1} c_n=0,
\end{equation}
where the new coefficients are given by
\begin{equation}
	\begin{split}
		\zeta''_0&=\zeta'_0;\\
		\beta''_n & =\beta'_n,\quad n<4;\\
		\gamma''_n & =\gamma'_n,\quad n<3;\\
		\delta''_n & =\delta'_n,\quad n<2;\\
	\end{split}
\end{equation}
and
\begin{subequations}
	\begin{equation}
		\alpha''_n=\alpha';
	\end{equation}
	\begin{equation}
		\beta''_n=\beta'_n-\frac{\eta'_{n-4}}{\zeta''_{n-4}}\alpha''_{n-1};
	\end{equation}
	\begin{equation}
		\gamma''_n=\gamma'_n-\frac{\eta'_{n-3}}{\zeta''_{n-3}}\beta''_n;
	\end{equation}
	\begin{equation}
		\delta''_n=\delta'_n-\frac{\eta'_{n-2}}{\zeta''_{n-2}}\gamma''_n;
	\end{equation}
	\begin{equation}
		\zeta''_n=\zeta'_n-\frac{\eta'_{n-1}}{\zeta''_{n-1}}\delta''_{n}.
	\end{equation}
\end{subequations}

The triple GE $5\to 4$ gives the four-term recurrence relation,
\begin{equation}
	\delta^{(3)}_{n-3} c_{n-3}+\gamma^{(3)}_{n-2} c_{n-2}+\beta^{(3)}_{n-1} c_{n-1}+\alpha^{(3)}_{n-1} c_n=0,
\end{equation}
with
\begin{equation}
	\begin{split}
		\delta^{(3)}_0&=\delta''_0;\\
		\beta^{(3)}_n & =\beta''_n,\quad n<3;\\
		\gamma^{(3)}_n & =\gamma''_n,\quad n<2;\\
	\end{split}
\end{equation}
and
\begin{subequations}
	\begin{equation}
		\alpha^{(3)}_n=\alpha''_n;
	\end{equation}
	\begin{equation}
		\beta^{(3)}_n=\beta''_n-\frac{\zeta''_{n-3}}{\delta^{(3)}_{n-3}}\alpha_{n-1}^{(3)},\quad n\ge 3;
	\end{equation}
	\begin{equation}
		\gamma^{(3)}_n=\gamma''_n-\frac{\zeta''_{n-2}}{\delta^{(3)}_{n-2}}\beta_n^{(3)},\quad n\ge 2;
	\end{equation}
	\begin{equation}
		\delta^{(3)}_n=\delta''_n-\frac{\zeta''_{n-1}}{\delta^{(3)}_{n-1}}\gamma_n^{(3)},\quad n\ge 1.
	\end{equation}
\end{subequations}

Finally, after the quadruple GE $4\to 3$, we obtain the three-term recurrence relation,
\begin{equation}
	\gamma^{(4)}_{n-2} c_{n-2}+\beta^{(4)}_{n-1} c_{n-1}+\alpha^{(4)}_{n-1} c_n=0,
\end{equation}
with
\begin{equation}
	\begin{split}
		\gamma^{(4)}_0 & =\gamma^{(3)}_0;\\
		\beta^{(4)}_n & =\beta^{(3)}_n,\quad n<2;\\
	\end{split}
\end{equation}
and
\begin{subequations}
	\begin{equation}
		\alpha^{(4)}_n=\alpha^{(3)}_n;
	\end{equation}
	\begin{equation}
		\beta^{(4)}_n=\beta^{(3)}_n-\frac{\delta_{n-2}^{(3)}}{\gamma_{n-2}^{(4)}}\alpha^{(4)}_{n-1},\quad n\ge 2;
	\end{equation}
	\begin{equation}
		\gamma^{(4)}_n=\gamma^{(3)}_n-\frac{\delta_{n-1}^{(3)}}{\gamma_{n-1}^{(4)}}\beta^{(4)}_{n},\quad n\ge 1.
	\end{equation}
\end{subequations}

\bibliography{references}
\end{document}